\documentclass[twocolumn,pre,showpacs]{revtex4} 

\usepackage{amssymb}
\usepackage{amsmath}
\usepackage{graphicx}
\usepackage{bm}
\renewcommand{\vec}[1]{\bm{#1}}

\begin{document}
\title{Biologically inspired learning in a layered neural net}
\author{J. Bedaux}
\email{bedaux@science.uva.nl}
\homepage{http://jasper.bedaux.net/}
\author{W.A. van Leeuwen}
\affiliation{Institute for Theoretical Physics, University of Amsterdam,
Valckenierstraat 65, 1018 XE Amsterdam, The Netherlands}

\date{\today}

\begin{abstract}
A feed--forward neural net with adaptable synaptic weights and fixed, zero or 
non--zero threshold potentials is studied, in the presence of a global feedback 
signal that can only have two values, depending on whether the output of the 
network in reaction to its input is right or wrong.

It is found, on the basis of four biologically motivated assumptions, that only 
two forms of learning are possible, Hebbian and Anti--Hebbian learning. 
Hebbian learning should take place when the output is right, while 
there should be Anti--Hebbian learning when the output is wrong.

For the Anti--Hebbian part of the learning rule a particular choice is made, 
which guarantees an adequate average neuronal activity without the need of 
introducing, by hand, control mechanisms like extremal dynamics. A network with 
realistic, i.e., 
non--zero threshold potentials is shown to perform its task of realizing the 
desired input--output relations best if it is sufficiently diluted, i.e.\ if 
only a relatively low fraction of all possible synaptic connections is realized.
\end{abstract}



\maketitle

\section{\label{introduction}Introduction}
In this article we will try to contribute to the study of biological neural 
networks, in particular with respect to learning. Since we are interested in the 
basic principles rather than subtle biological details, we will use simplified 
models, although we do not allow for any properties that are unrealistic from a 
biological point of view. In this way we hope to reveal the essentials, without 
blurring the analysis with (irrelevant, biological) details.

It is also not our aim to construct a network that is optimized for some particular 
task; our only purpose is to study a model that resembles an actual biological 
neural net and the way it might learn.

In section \ref{model} we define the model that we will study: a simple 
feed--forward network with one hidden layer of which not all possible connections 
are present (i.e., dilution unequal zero). Each neuron of the net has three 
variables associated with it: a (fixed) threshold potential $\theta$, a 
(variable) membrane potential $h$ and a state $x$, which is assumed to take two 
values only, depending on the fact whether the neuron fires or is quiescent.

This article fits into a line of biologically motivated research. Chialvo and 
Bak \cite{bak} suggested learning by punishment only, via the release of some 
hormone. Heerema et al.\ \cite{leeuwen} derived rules for a biological neural 
net when it acts as a memory. Bosman et al.\ \cite{bosman} added this rule to 
the model of Chialvo and Bak, and, thereby, found a significant improvement of 
the network's performance.

The dynamics of a neural net is determined by a rule that tells a neuron when to 
fire. A biological neuron fires if its membrane potential $h$ exceeds its 
threshold $\theta$. In the model of Chialvo and Bak this biological rule is 
replaced by the rule that in each layer a fixed number of neurons, having the 
highest membrane potentials, fire. They refer to this rule by the name of 
`extremal dynamics'. Extremal dynamics has the drawback that the number of 
active neurons is artificially fixed, restricting the number of possible output 
states. Alstr{\o}m and Stassinopoulos \cite{alstrom} did not use 
extremal dynamics, but they ran into the difficulty that the network's activity 
became too small or too large for the network to function satisfactorily. In 
order to keep the activity at a desired level, they adapted the neuron threshold 
potentials $\theta$ at each step of the learning process. 

It is one of our purposes to find a way in which a biological net can keep its 
neuronal activity around an acceptable level. Instead of putting in, by hand, 
some controlling mechanism, we 
started from four biologically motivated assumptions (\ref{hebbvsantihebb}),
which lead to the conclusion that only the so--called Hebbian and 
Anti--Hebbian learning rules are the plausible ones. We
show that the Hebbian learning rule fixes and strengthens the action of the
network at the moment it is applied, while the Anti--Hebbian learning rule 
does the opposite, and, when applied repeatedly, will change 
the network's action. We conclude that Hebbian learning should be associated 
with reward, and should be applied if the network realizes the desired output 
state in reaction to its input, while Anti--Hebbian learning should be 
associated with punishment, and should be applied when the output of the network 
is wrong, in order to enable the network to search for better output (\ref{reinf}).

In section \ref{explicitrules} we propose a particularly simple Anti--Hebbian
learning rule (essentially two constants) of which we expect, however,
that it will keep the activity of the neural network around a desired
value automatically, i.e., without the need of some controlling mechanism.
In section \ref{num1}, we perform a number of numerical simulations, 
the details of which are given in section \ref{num0}, in order 
to verify whether our rule is
capable of keeping the activity around a desired level. Because the Hebbian 
learning rule is already studied in \cite{leeuwen}, we will first focus, in 
section \ref{num1}, on the effect of the Anti--Hebbian learning rule. Our 
simulations show that our Anti--Hebbian learning rule indeed 
is successful in keeping 
the average activity around a desired level, and, moreover,
is very efficient in generating 
different output states with adequate activities.

Having studied the effect of the Anti--Hebbian learning rule we simulate, in 
section \ref{num2}, networks using the complete learning rule, including both 
the Hebbian and the Anti--Hebbian contributions. 
We show that the complete learning rule 
enables the network to learn a number of input--output relations with an 
acceptable efficiency. Furthermore, it is shown that for non--zero 
threshold potentials, the performance is only acceptable if the network is 
sufficiently diluted. So `cutting away' synaptic connections enhances the 
performance of the network. The observation that -- in our model -- some degree
of non--connectedness is a conditio sine qua non for a properly functioning
biological net, is one of our main observations.

The article closes with conclusions (section \ref{conclusions}) and an outlook 
(section \ref{outlook}).

\section{\label{model}Modeling learning}
In this section we describe our model and define quantities like activity and 
performance.

\subsection{\label{network}Network model}
\begin{figure}
\includegraphics*[width=\columnwidth]{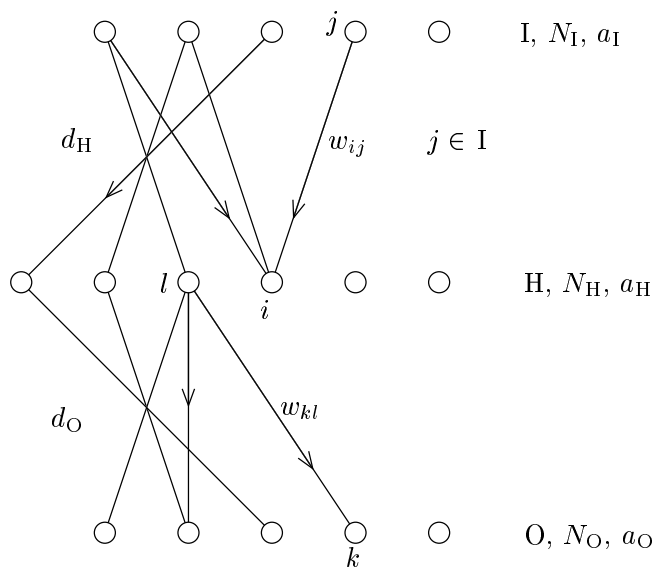}
\caption{The feed--forward network has one hidden layer. The symbols I, H and O 
refer to the Input, Hidden and Output layers. In order to denote that a neuron 
$i$ of the net belongs to one of these layers we will write $i \in \text{I}$, $i 
\in \text{H}$ or $i \in \text{O}$, respectively. Furthermore, $N_\text{X}$ 
denotes the number of neurons in layer X (X = I, H or O). 
The weights of the synapses connecting either the input layer 
with the hidden layer or the hidden layer with the output layer both are 
indicated by the symbol $w_{ij}$. In this feed--forward network there are 
maximally $N_\text{I}N_\text{H}$ possible connections from the input layer to 
the hidden layer. If only the fraction $(1 - d_\text{H})$, $0 \leq d_\text{H} 
\leq 1$, of these connections to the hidden layer is actually realized, we call 
$d_\text{H}$ the dilution (of the connections with H). Similarly, $d_\text{O}$ 
is the dilution of the connections from the hidden layer to the output layer.}
\label{fig1}
\end{figure}
We consider a feed--forward neural net with one hidden layer (figure 
\ref{fig1}). The net is taken to be diluted, i.e.\ not all neurons are connected 
to all others. There are connections from the input layer to the hidden layer 
and from the hidden layer to the output layer but there are no direct 
connections from the input layer to the output layer nor any lateral or 
feed--back connections.

Let us suppose that the neural net consists of $N$ binary neurons $i$, $i = 1, 
2, \ldots, N$. We will use the symbols I, H and O to 
refer to the Input, Hidden and Output layers. In order to denote that a neuron 
$i$ of the net belongs to one of these layers we will write $i \in \text{I}$, $i 
\in \text{H}$ or $i \in \text{O}$, respectively. The state $x_i$ of neuron $i$ 
either is active ($x_i = 1$) or is non--active ($x_i = 0$). The potential $h_i$, 
the difference in potential between the inner and the outer part of a neuron, is 
supposed to depend linearly on the activities of the incoming synaptic 
connections:
\begin{equation}\label{potential}
h_i = \sum_{j \in V_i} w_{ij} x_j
\end{equation}
where $V_i$ is the collection of neurons $j$ that have an afferent synaptic 
connection to neuron $i$. This formula can be viewed as the defining expression 
for the weights $w_{ij}$. Since the $x_i$'s are chosen to be dimensionless, the 
weights have the dimension of a potential. Let $\theta_i$ be the potential that 
should be exceeded in order that neuron $i$ becomes active, i.e.,
\begin{equation}\label{state}
x_i = 1 \quad \text{if} \quad h_i > \theta_i, \quad
x_i = 0 \quad \text{if} \quad h_i < \theta_i.
\end{equation}
An alternative way to specify the state of neuron $i$ as given by equation 
(\ref{state}), is to write
\begin{equation}\label{dynamics}
x_i = \Theta_\text{H}(h_i - \theta_i).
\end{equation}
The function $\Theta_\text{H}$ is the Heaviside step--function 
[$\Theta_\text{H}(y) = 0$ if $y < 0$ and $\Theta_\text{H}(y) = 1$ if $y > 
0$]. The state of the neurons of the input layer determine the state of the 
hidden layer via (\ref{potential}) and (\ref{dynamics}). The state of the hidden 
layer, in turn, determines the state of the output layer. Thus, if the states 
$x_i$ for $i \in \text{I}$ are given at each time step $t_n$, we can determine 
the network state at every time $t_n$, once we have a rule that fixes the 
$w_{ij}$ at time $t_n$ ($n = 0,1,2,\ldots$).

The number of neurons in layer X (X = I, H or O) is denoted by $N_\text{X}$. 
The activity $a_\text{X}$ in layer X is defined by 
\begin{equation}\label{activity}
a_\text{X} = \frac{1}{N_\text{X}} \sum_{i \in \text{X}}x_i, \quad \text{X} = 
\text{I}, \text{H}, \text{O}.
\end{equation} The weights of the synapses connecting either the input layer 
with the hidden layer or the hidden layer with the output layer both are 
indicated by the symbol $w_{ij}$. 

In this feed--forward network there are maximally $N_\text{I}N_\text{H}$ 
possible connections from the input layer to the hidden layer. If only the 
fraction $(1 - d_\text{H})$, $0 \leq d_\text{H} \leq 1$, of these connections to 
the hidden layer is actually realized, we call $d_\text{H}$ the dilution (of the 
connections with H). Similarly, $d_\text{O}$ is the dilution of the connections 
from the hidden layer to the output layer.

\subsection{\label{iorelations}Learning input--output relations}
Let us denote the state of layer X by $\vec{x}_\text{X} = (x_{\text{X}1}, 
x_{\text{X}2}, \ldots, x_{\text{X}N_\text{X}})$. We want the network to 
associate with a particular, prescribed input state $\vec{\xi}^\mu_\text{I}$, 
chosen from a collection of $p$ input states $\vec{\xi}^\mu_\text{I}$ ($\mu = 1, 
\ldots, p$), the particular state $\vec{\xi}^\mu_\text{O}$, chosen from a 
collection of $p$ prescribed output states $\vec{\xi}^\mu_\text{O}$ ($\mu = 1, 
\ldots, p$). The goal of the learning process is that the network will be able 
to generate, for \emph{all} input patterns $\vec{\xi}^\mu_\text{I}$, the correct 
output pattern $\vec{\xi}^\mu_\text{O}$. This will be achieved by a learning 
procedure, in which the weights are adapted, stepwise, according to some rule.

If we present a pattern $\vec{\xi}^\mu_\text{I}$ to the network by setting 
$\vec{x}_\text{I}$ equal to $\vec{\xi}^\mu_\text{I}$, the network will respond 
by generating an output state $\vec{x}_\text{O}$ which, in general, will not be 
equal to the desired output state $\vec{\xi}^\mu_\text{O}$. We will associate a 
variable $r$ with each of the following two possibilities: $r = 1$ if the output 
is right, i.e., if $\vec{x}_\text{O} = \vec{\xi}^\mu_\text{O}$ and $r = 0$ if 
the output is wrong, i.e., if $\vec{x}_\text{O} \neq \vec{\xi}^\mu_\text{O}$. In 
practice, we proceed as follows. We present the first input pattern 
$\vec{\xi}^1_\text{I}$ to the net, which results in an output state 
$\vec{x}_\text{O}$. We keep repeating this until $\vec{x}_\text{O}$ equals 
$\vec{\xi}^1_\text{O}$, the output pattern to be associated with the input 
pattern $\vec{\xi}^1_\text{I}$. The net then has `learned' the first input--output 
relation $(\vec{\xi}^1_\text{I}, \vec{\xi}^1_\text{O})$. Next, we present 
the second input pattern, and continue until the second input--output relation 
$(\vec{\xi}^2_\text{I}, \vec{\xi}^2_\text{O})$ has been learned, and so on until 
the $p$--th input--output relation $(\vec{\xi}^p_\text{I}, 
\vec{\xi}^p_\text{O})$ has been learned. While doing so, we continuously change 
the weights according to some $r$--dependent rule: the `learning rule'.

When an input--output relation has been learned, one or more input--output 
relations learned earlier may have been forgotten. Therefore, we start a 
second round, in which we try to learn all $p$ input--output relations 
$(\vec{\xi}^\mu_\text{I}, \vec{\xi}^\mu_\text{O})$ again. In order to prevent 
possible effects due to a specific learning order, the patterns are presented to 
the network in random order: each round their order is shuffled. After a number 
of rounds all input--output relations should be recalled at once, i.e., input 
$\vec{\xi}^\mu_\text{I}$ should result immediately in output 
$\vec{\xi}^\mu_\text{O}$ for all input--output relations.

Do learning rules exist which accomplish all this? The answer is positive 
\cite{bak, bosman}. A particular useful one is the one proposed in section 
\ref{seclearning}.

\subsection{\label{perf}Performance}
In order to judge the performance of our network, we need some measure.

The a priori probability $P(N_\text{O}^{(a)})$ that the state of the 
output layer $\vec{x}_\text{O}$ equals $(1, 1, \ldots, 1, 0, 0, \ldots,0)$, 
i.e., the state with the first $N_\text{O}^{(a)}$ elements equal to one and the 
remaining last $N_\text{O} - N_\text{O}^{(a)}$ elements equal to zero, is given 
by
\begin{equation}\label{pee}
P(N_\text{O}^{(a)}) = p_\text{O}^{N^{(a)}_\text{O}}
(1 - p_\text{O})^{N_\text{O} - N^{(a)}_\text{O}}
\end{equation}
where $p_\text{O}$ is the probability that an arbitrary neuron of the output 
layer is active ($0 \leq p_\text{O} \leq 1$). Note that the probability to find 
an output pattern with $N_\text{O}^{(a)}$ active neurons at arbitrary places is 
also given by (\ref{pee}). Let $N^{(a)}_\text{O}(\vec{\xi}^\mu_\text{O})$ be the 
number of active neurons in pattern $\vec{\xi}^\mu_\text{O}$. Then, the average 
a priori number of trials needed to arrive at the desired output pattern equals 
$1/P(N^{(a)}_\text{O}(\vec{\xi}^\mu_\text{O}))$. If $p$ input--output 
relations $(\vec{\xi}^\mu_\text{I}, \vec{\xi}^\mu_\text{O})$ are to be realized, 
($\mu = 1, \ldots, p$), the average a priori number $M_a$ of trials needed 
equals
\begin{equation}\label{Ma}
M_a = \sum_{\mu = 1}^p 1/P(N^{(a)}_\text{O}(\vec{\xi}^\mu_\text{O})).
\end{equation}
Comparing the average a priori number of trials and the actual 
average number of trials
$M$ needed to learn all $p$ input--output relations we get a 
measure of how well the network performs. This leads us to define the 
performance $R$ as the quotient
\begin{equation}\label{performance}
R := M_a/M
\end{equation}
with $M_a$ given by (\ref{Ma}). Note that the performance tends to zero when the 
network is unable to learn all input--output relations ($M \rightarrow \infty$) 
and the performance will be 1 for an `ideal' network ($M = M_a$).

In some models \cite{bak, bosman}, the number of neurons that fire 
is fixed in some way. The a priori probability 
$P_\text{fixed}(N_\text{O}^{(a)}(\vec{\xi}^\mu_\text{O}))$ that the 
output state $\vec{x}_\text{O}$ equals the output pattern 
$\vec{\xi}^\mu_\text{O}$ is then given by 
\begin{equation}\label{pextr}
P_\text{fixed}(N_\text{O}^{(a)}(\vec{\xi}^\mu_\text{O})) = 
\frac{N_\text{O}^{(a)}(\vec{\xi}^\mu_\text{O})!(N_\text{O} - N_\text{O}^{(a)}
  (\vec{\xi}^\mu_\text{O}))!}{N_\text{O}!},
\end{equation}
the inverse of the number of ways that a state with $N_\text{O}^{(a)}$ active 
neurons can be realized.

\subsection{\label{dista}Activity distribution}
Let us consider an arbitrary 
collection of $M$ neurons, all of which have (independent) probabilities $p$ to 
fire ($0 \leq p \leq 1$). Then the probability $Q(a)$ that the activity of this 
collection of $M$ neurons equals $a$ is given by
\begin{equation}\label{Qa}
Q(a) = \frac{M!}{(Ma)![M(1 - a)]!}p^{Ma}(1 - p)^{M(1 - a)},
\end{equation}
where $Ma$ is the number of neurons that fire ($Ma$ is an integer). 
This formula differs from (\ref{pee}) by a binomial factor, corresponding
to the fact that there are $M!/\{(Ma)![M(1 - a)]!\}$ possible
states with the same activity $a$.
We will use this formula in section \ref{num1} when we compare $Q(a)$ to the
actual activity of the net.

\subsection{\label{stabcoef}Stability coefficient}
Let us define the `two time stability coefficient'
\begin{equation}\label{gamma}
\gamma_i(t_n, t_m) := \left(2x_i(t_n) - 1\right)\left(h_i(t_m) - \theta_i\right)
\end{equation}
which is a generalization of the usual stability coefficient \cite{kinzel}. 
With equation (\ref{dynamics}) it follows that, at two times $t_n$ and $t_m$ we 
have
\begin{equation}\label{gammaposnm}
\gamma_i(t_n, t_m) > 0 \Longrightarrow x_i(t_n) = x_i(t_m)
\end{equation}
\begin{equation}\label{gammanegnm}
\gamma_i(t_n, t_m) < 0 \Longrightarrow x_i(t_n) \neq x_i(t_m).
\end{equation}
In particular, if $n = m$, we have
\begin{equation}\label{gammaposnn}
\gamma_i(t_n, t_n) > 0.
\end{equation}
The state of neuron $i$ does not change if $\gamma_i(t_n, t_m)$ is positive. 
Moreover, the larger $\gamma_i(t_n, t_n)$, the more stable the system is with 
respect to changes in the synaptic weights \cite{krauth, gardner}. 
The $\gamma$'s will play a role in section \ref{reinf}.

\section{\label{seclearning}Determining the learning rule} 
As noted in the introduction, many neural networks have a tendency to develop into 
a state where most neurons are either all active or non--active. Chialvo and Bak 
\cite{bak} and Alstr\o m and Stassinopoulos \cite{alstrom} solved this problem 
in a biologically less plausible way: the authors do not consider the 
threshold potentials at all \cite{bak}, or allow them to grow indefinitely 
\cite{alstrom}. We approach the problem in a different way: we start from four 
biologically plausible restrictions on the learning rule. This leads to the 
conclusion that the learning rule can be seen as a superposition of two types of 
learning, Hebbian and Anti--Hebbian learning. Taking for the Anti--Hebbian part 
the most simple ansatz one can think of in dependence of the state $x_i$ of the 
postsynaptic neuron $i$, we arrive at a learning rule that, a priori, is 
biologically plausible. Moreover, it turns out that the learning rule found in
this way is such that the neuronal activity is well--behaved.

\subsection{\label{hebbvsantihebb}Biologically plausible learning rules}
Let us suppose that the weights $w_{ij}(t_n)$ may be changed stepwise by some 
learning process
\begin{equation}\label{wij}
w_{ij}(t_{n+1}) = w_{ij}(t_n) + \Delta w_{ij}(t_n)
\end{equation}
where $j$ is the presynaptic and $i$ the postsynaptic neuron. The changes,
given by $\Delta w_{ij}(t_n)$, are the quantities we are after.

We start from four assumptions, each of which needs not be strictly true, but 
each of which is very plausible biologically, at least in first approximation.
The contents of these assumptions generalize the assumptions of Heerema and 
Van Leeuwen \cite{leeuwen} by including the effect of the feed--back signal $r$.

\begin{itemize}
\item[\emph{i.}] The changes in $w_{ij}$, $\Delta w_{ij}$, depend on the global 
variable $r$ and the local variables $x_i$, $x_j$, $h_i$, $\theta_i$ and 
$w_{ij}$, i.e.,
\begin{equation}\label{j1}
\Delta w_{ij} = \Delta w_{ij}(r, x_i, x_j, h_i, \theta_i, w_{ij}).
\end{equation}

Biologically this means that only variables that can be `felt' at the synapse 
between $j$ and $i$ can influence the weight change. Of course, this includes 
the global variable $r$, which could be realized in an actual biological system 
by some chemical substance released throughout the brain, and the strength of 
the weight $w_{ij}$ itself. Since the state $x_j$ of neuron $j$ determines 
whether or not neurotransmitters are released at the synapse, $x_j$ can influence 
the weight change. Because the synapse is located at the dendrites or cell body 
of neuron $i$, we suppose that variables local to neuron $i$, the state $x_i$ 
and the potentials $h_i$ and $\theta_i$, also can influence the strength of 
the weight change.
\item[\emph{ii.}] The \emph{sign} of $\Delta w_{ij}$ depends on $r$, and the 
neuron states $x_i$ and $x_j$ \emph{only}, i.e.,
\begin{equation}\label{j2}
\Delta w_{ij} = \sigma(r, x_i, x_j)\varepsilon_{ij}(r, x_i, x_j, 
h_i, \theta_i, w_{ij})
\end{equation}
where $\sigma$ equals $-1$ or $+1$ and where $\varepsilon_{ij} \geq 0$.

Biologically this means that as long as the states of the presynaptic and 
postsynaptic neurons do not change \emph{and} the global feed--back signal $r$ 
does not change, the sign of the weight change will not change, i.e., the 
variables $h_i$, $\theta_i$ and $w_{ij}$ can influence the \emph{magnitude} of 
the weight change, but they cannot switch the learning from an increase to a 
decrease as long as the states and $r$ do not change. \item[\emph{iii.}] There 
is only a change in $w_{ij}$ if the presynaptic neuron $x_j$ is active, i.e.,
\begin{equation}\label{j3}
\varepsilon_{ij} = \varepsilon_{ij}(r, x_i, h_i, \theta_i, w_{ij}) x_j.
\end{equation}

Biologically this means that if the presynaptic neuron $j$ 
does not fire, there will 
be no change in the synaptic efficacy $w_{ij}$. This is supposed because we 
think that it is unlikely that the synaptic efficacy will change if nothing 
happens, i.e., no neurotransmitters are released into the synaptic gap. 
Substituting (\ref{j3}) into (\ref{j2}) we find
\begin{equation}\label{j3a}
\Delta w_{ij} = \sigma(r, x_i)\varepsilon_{ij}(r, x_i, h_i, \theta_i, w_{ij}) 
x_j,
\end{equation}
where we replaced $\sigma(r, x_i, x_j)$ by $\sigma(r, x_i)$, which is allowed
since $\Delta w_{ij}$ is only non--zero if $x_j$ equals $1$.
\item[\emph{iv.}] Both in case $r = 0$ and $r = 1$, there is not only 
enhancement or only diminishment of the weights. This implies that $\sigma(r = 
0, x_i)$ must take on both the values $+1$ and $-1$.
The same is true 
for $\sigma(r = 1, x_i)$. Since $x_i$ is the only variable that is still 
available to influence $\sigma$, we must have $\sigma(r, x_i = 0) = 
-\sigma(r, x_i = 1)$, or, equivalently
\begin{equation}\label{w4}
\sigma(r, x_i) = \sigma(r)(2x_i - 1)
\end{equation}
where $\sigma(r)$ equals $-1$ or $+1$.

Biologically, this means that we think that it is implausible that weights can 
either only increase or only decrease. If the weights would only increase, all 
the membrane potentials (\ref{potential})
would only increase, and all neurons would end up 
firing [see eq.\ (\ref{dynamics})]. 
In case the weights would only decrease, all membrane potentials would 
only decrease, and all neurons would become non--active. Since $r = 0$ can be 
true for long times as long as something is not learned, and $r = 1$ can also 
last for long times as long as the behavior is as desired, the assumption must hold 
true both for $r = 0$ and $r = 1$.
\end{itemize}
Inserting (\ref{w4}) into (\ref{j3a}) we finally have
\begin{equation}\label{j4}
\Delta w_{ij} = \sigma(r)\varepsilon_{ij}(r, x_i, h_i, \theta_i, 
w_{ij})(2x_i - 1)x_j,
\end{equation}
where $\sigma(r) = \pm 1$ (where $r$ can take on the values $0$ and $1$) 
and $\varepsilon_{ij} \geq 0$.

\subsection{\label{reinf}Implementing reward and punishment}
We will now discuss the effects of the weight changes (\ref{j4}) for
$\sigma(r) = +1$ and $\sigma(r) = -1$.
To that end, consider neuron $i$, a fixed but 
arbitrary neuron of the network. Suppose that the neurons $j \in V_i$ do not 
change their states during the time step $t_n \rightarrow t_{n + 1}$:
\begin{equation}\label{xjtn}
x_j(t_{n + 1}) = x_j(t_n),\quad j \in V_i
\end{equation}
Now, we multiply both sides of (\ref{wij}) by $x_j(t_n)$, sum over all indices 
$j \in V_i$, and subtract the threshold potential $\theta_i$ to obtain
\begin{equation}\label{hitnp1}
h_i(t_{n + 1}) - \theta_i = h_i(t_n) - \theta_i + \sum_{j \in V_i} 
\Delta w_{ij}(t_n)x_j(t_n)
\end{equation}
where we used (\ref{potential}) and (\ref{xjtn}). Next, we multiply by the 
factor $(2x_i(t_n) - 1)$. Using equation (\ref{gamma}), we then get
\begin{equation}\label{gitn}
\gamma_i(t_n, t_{n + 1}) = \gamma_i(t_n, t_n) + \sum_{j \in V_i} \Delta 
w_{ij}(t_n) (2x_i(t_n) - 1) x_j(t_n).
\end{equation}
We now distinguish between $\sigma(r) = +1$ and $\sigma(r) = -1$.

Substituting (\ref{j4}) into (\ref{gitn}) with $\sigma(r) = +1$ we find, 
using $(2x_i - 1)^2 = 1$ and $x_j^2 = x_j$, that 
\begin{equation}\label{gitnp1}
\gamma_i(t_n, t_{n + 1}) = \gamma_i(t_n, t_n) + \sum_{j \in V_i} 
\varepsilon_{ij}x_j(t_n).
\end{equation}
Recalling that $\gamma_i(t_n, t_n)$ and $\varepsilon_{ij}$ are 
positive, we see that $\gamma_i(t_n, t_{n + 1})$ is positive. Hence, according 
to (\ref{gammaposnm}) the state of neuron $i$ has not changed: 
\[ x_i(t_{n + 1}) = x_i(t_n). \]
In our case of a feed--forward network, the neurons of the hidden layer get 
their input from neurons of the input layer only. Consequently, if the input 
layer does not change, going from time $t_n$ to time $t_{n + 1}$, the state of a 
neuron $i$ of the hidden layer does not change. This holds true for any neuron 
$i$ of the hidden layer, implying that the input of the output layer will not 
change either. 
In other words, the output of the net does not change when the input 
remains the same, although the weights $w_{ij}(t_n)$ change to $w_{ij}(t_{n + 
1})$ according to the rule (\ref{wij}). Hence, the rule (\ref{j4}) with 
$\sigma(r) = +1$ conserves an input--output relation.

Moreover, since $\gamma_i(t_n, t_{n + 1})$ is larger than $\gamma_i(t_n, t_n)$, 
as follows from (\ref{gitnp1}), the new net is more stable. In other words, 
learning with $\sigma(r) = +1$ 
engraves an input--output relation into the memory of the net 
by properly adapting its weights.

Next, substituting (\ref{j4}) into (\ref{gitn}), but now with $\sigma(r) = -1$,
we find
\begin{equation}\label{gitnp1A}
\gamma_i(t_n, t_{n + 1}) = \gamma_i(t_n, t_n) - \sum_{j \in V_i} 
\varepsilon_{ij}x_j(t_n).
\end{equation}
Since both $\gamma_i(t_n, t_n)$ and $\varepsilon_{ij}$ are positive, 
$\gamma_i(t_n, t_{n + 1})$ is smaller than $\gamma_i(t_n, t_n)$. Hence, 
the learning rule (\ref{j4}) with $\sigma(r) = -1$ 
has the effect of decreasing the stability of the 
network. As long as the state of the network does not change (i.e.\ $x_j
(t_{n + 1}) = x_j(t_n)$ for all $j$), all stability coefficients 
$\gamma_i(t_n, t_{n + 1})$ decrease, and, at a certain moment $t_m$ ($m > n$), 
$\gamma_i(t_m, t_{m + 1})$ will become negative, at least for one neuron $i$, 
implying, with (\ref{gammanegnm}), that \[ x_i(t_{m + 1}) \neq x_i(t_m). \] 
Consequently, repeated learning with $\sigma(r) = -1$ 
will result in a change of the output related to the same input.

We now come to the main conclusion of this section. If the network output is the 
wrong one ($r = 0$), we must adapt the weights such that other output results, 
i.e., we should use $\sigma(0) = -1$. 
If the network output is the right one ($r = 1$), we should use $\sigma(1) = +1$
to consolidate this situation. Hence, we may conclude
\begin{equation}\label{sigma}
\sigma(r) = 2 r - 1.
\end{equation}
With (\ref{sigma}) we may write instead of (\ref{j4})
\begin{equation}\label{j1wij}
\Delta w_{ij} = r \Delta w_{ij}^\text{H} + (1 - r) \Delta w_{ij}^\text{A}
\end{equation}
where
\begin{eqnarray}\label{H}
\Delta w_{ij}^\text{H} & = & +\varepsilon_{ij}^\text{H}(x_i, h_i, 
\theta_i, w_{ij})(2x_i - 1)x_j\\
\label{A}
\Delta w_{ij}^\text{A} & = & -\varepsilon_{ij}^\text{A}(x_i, h_i, 
\theta_i, w_{ij})(2x_i - 1)x_j
\end{eqnarray}
with
\begin{eqnarray}\label{eps1}
\varepsilon_{ij}^\text{H} & = & \varepsilon_{ij}(r = 1, x_i, h_i, 
\theta_i, w_{ij})\\
\label{eps2}
\varepsilon_{ij}^\text{A} & = & \varepsilon_{ij}(r = 0, x_i, h_i, 
\theta_i, w_{ij}).
\end{eqnarray}
If both the presynaptic and postsynaptic neurons are active ($x_j = x_i = 1$),
the terms $\Delta w_{ij}^\text{H}$ and $\Delta w_{ij}^\text{A}$ are positive
and negative, respectively. We will refer to them by the names of Hebbian
and Anti--Hebbian learning. In conclusion, we see from (\ref{j1wij}) that
reward ($r = 1$) and punishment ($r = 0$) may be associated in a unique way
with Hebbian
and Anti--Hebbian learning, respectively. A similar rule has been postulated
by Barto and Anandan \cite{barto}.

What remains is to find explicit expressions for the (positive) functions 
$\varepsilon_{ij}^\text{H}$ and $\varepsilon_{ij}^\text{A}$. This will be 
the subject of the next section.

\subsection{\label{explicitrules}Determining explicit rules}
For the Hebbian function $\varepsilon_{ij}^\text{H}$ we choose the biologically
plausible expression derived in \cite{leeuwen}:
\begin{equation}\label{epsilonijH}
\varepsilon_{ij}^\text{H} = \eta_i\left[\kappa - (h_i - \theta_i)(2x_i -
1)\right]
\end{equation}
where $\eta_i$ and $\kappa$ are constants. Note that $\kappa$ must be large 
enough in order that $\varepsilon_{ij}^\text{H}$ be positive.

We now come to the Anti--Hebbian function $\varepsilon_{ij}^\text{A}(x_i, h_i, 
\theta_i, w_{ij})$. Experimentally nor theoretically there are clues regarding
the precise form of this term. Therefore, we simply choose two (positive) 
constants, $c_i^{(1)}$ and $c_i^{(2)}$: $\varepsilon_{ij}^\text{A}(x_i = 0) = 
c_i^{(1)}$ and $\varepsilon_{ij}^\text{A}(x_i = 1) = c_i^{(2)}$. Since any two 
positive constants can be expressed as $c_i^{(1)} = \rho_i \alpha_i$ and 
$c_i^{(2)} = \rho_i (1 - \alpha_i)$ where $\rho_i$ and $\alpha_i$ are two 
other constants with $0 < \alpha_i < 1$ and $\rho_i > 0$, we can write 
\begin{equation}\label{epsilonijAprime}
\varepsilon_{ij}^\text{A}(x_i = 0) = \rho_i \alpha_i, \quad 
\varepsilon_{ij}^\text{A}(x_i = 1) = \rho_i (1 - \alpha_i)
\end{equation}
or, equivalently,
\begin{equation}\label{epsilonijA}
\varepsilon_{ij}^\text{A} = \rho_i \left[{\textstyle\frac{1}{2}} - 
(\alpha_i - {\textstyle\frac{1}{2}})(2 x_i - 1)\right].
\end{equation} Note that the expression (\ref{epsilonijA}) can be obtained 
from (\ref{epsilonijH}) by the substitutions $\eta_i \rightarrow \rho_i$, $h_i 
\rightarrow \alpha_i$, $\kappa \rightarrow \frac{1}{2}$ and $\theta_i 
\rightarrow \frac{1}{2}$. However, for (\ref{epsilonijH}) exists a derivation, 
whereas (\ref{epsilonijA}) is an educated guess only.

Upon substituting (\ref{epsilonijH}) and (\ref{epsilonijA}) into (\ref{H}) and 
(\ref{A}) we obtain
\begin{eqnarray}\label{deltawijH2}
\Delta w_{ij}^\text{H} & = & \eta_i\left[\kappa(2x_i - 1) - (h_i -
\theta_i)\right]x_j\\
\label{deltawijA2}
\Delta w_{ij}^\text{A} & = & -\rho_i(x_i - \alpha_i)x_j.
\end{eqnarray}

The numerical study of section \ref{num1} will show that the effect of the 
Anti--Hebbian term (\ref{deltawijA2}) is that the average neuronal activity 
of the network takes a value controlled by the parameter $\alpha$.

\section{\label{num0}Simulating the network}
This section is intended mainly for readers interested in technical details of 
the simulations.

\subsection{\label{parms}Determining parameter values}
Up to now we did not specify the parameters $\eta_i$, $\kappa$, $\theta_i$, 
$\rho_i$ and $\alpha_i$ occurring in the learning rule (\ref{j1wij}) combined 
with (\ref{deltawijH2}) and (\ref{deltawijA2}). In \cite{leeuwen} it is argued 
that the Hebbian learning rate $\eta_i$ should be proportional to the inverse of 
the average number of neurons $j \in V_i$ that fire. It is argued also that a 
reasonable approximation will suffice. Therefore, we may choose $\eta_i$ equal 
to $\eta_\text{X}$ for all $i \in \text{X}$, i.e., we may choose $\eta_i$ the 
same for all neurons $i$ of layer X ($\text{X} = \text{H}, \text{O}$):
\begin{equation}\label{etaHO1}
\eta_i = \eta_\text{H}, \quad i \in \text{H}, \quad
\eta_i = \eta_\text{O}, \quad i \in \text{O}
\end{equation}
with
\begin{equation}\label{etaHO2}
\eta_\text{H} = \eta\frac{1}{\overline{a_\text{I}}N_\text{I}(1 - d_\text{H})}, \quad
\eta_\text{O} = \eta\frac{1}{\overline{a_\text{H}}N_\text{H}(1 - d_\text{O})}
\end{equation}
where the bar denotes a time average and where $\eta$ is some positive constant, 
which we will call the (global) learning rate. Note that $\eta_\text{H}$ is the 
learning rate associated with the connections from the input layer I to the 
hidden layer H. Similarly, $\eta_\text{O}$ is associated with connections from H 
to O.

Analogously we take for the Anti--Hebbian learning rate, or punishment rate,
$\rho$:
\begin{equation}\label{rhoHO1}
\rho_i = \rho_\text{H}, \quad i \in \text{H}, \quad
\rho_i = \rho_\text{O}, \quad i \in \text{O} 
\end{equation}
with
\begin{equation}\label{rhoHO2}
\rho_\text{H} = \rho\frac{1}{N_\text{I}\overline{a_\text{I}}(1 - d_\text{H})}, \quad
\rho_\text{O} = \rho\frac{1}{N_\text{H}\overline{a_\text{H}}(1 - d_\text{O})}.
\end{equation}

We will set the parameters $\theta_i = \theta_\text{X}$, $i \in \text{X}$ and 
$\alpha_i = \alpha_\text{X}$, $i \in \text{X}$, i.e., we take these parameters 
the same also for neurons belonging to one and the same layer.

The margin parameter $\kappa$ will be fixed at the value 1, in agreement with 
literature (see e.g.\ \cite{leeuwen}). This can be done, because, instead of 
varying $\kappa$, one can also vary both the learning and punishment rates 
$\eta$ and $\rho$, with the same effect.

\subsection{\label{noise}Addition of noise}
Since the equations describing the network dynamics are deterministic, and the 
number of possible states of the network is finite, the system may suffer from 
periodic behavior, which is not realistic biologically, since in an actual 
biological net there is always some disturbing effect. Therefore, when 
performing our simulations, we add some noise, in order to mimic reality.

To that end we employ the Gaussian distribution
\begin{equation}\label{Fx}
F(x, \mu, \sigma) = \frac{1}{\sigma \sqrt{2 \pi}}e^{-(x - \mu)^2 / 2 \sigma^2}
\end{equation}
with mean $\mu$ and standard deviation $\sigma$. In our simulations we will 
replace the deterministic value $\Delta w_{ij}$ by the probabilistic value 
$\Delta w^{(\text{noisy})}_{ij}$, the distribution of which is given by
\begin{equation}\label{Px}
P(\Delta w^{(\text{noisy})}_{ij}) = F(\Delta w^{(\text{noisy})}_{ij}, \mu, 
\sigma)
\end{equation}
with
\begin{equation}\label{musigma}
\mu = \Delta w_{ij}, \quad \sigma = \Delta w_{ij} \delta
\end{equation}
i.e., we replaced $\Delta w_{ij}$ by a Gaussian distributed quantity $\Delta 
w^{(\text{noisy})}_{ij}$ with mean value $\Delta w_{ij}$ and standard deviation 
$\Delta w_{ij}\delta$. Note that we have chosen the standard deviation of the 
noise proportional to $\Delta w_{ij}$ with proportionality constant $\delta$.

\subsection{\label{secinit}Initializing the network}
As long as a network does not memorize input--output relations, the Anti--Hebbian 
term of the learning rule will adapt the weights in such a way that all 
neurons will `hesitate' between firing and not firing. When the Anti--Hebbian 
learning rule is applied for a long time, the network can be viewed as `fresh': 
the network has not stored any information and can quickly change its behavior. 
We want to start our simulations with networks that have their weights 
distributed according to such a fresh distribution. This can be accomplished by 
starting with arbitrary weights and applying the Anti--Hebbian learning rule a 
large number of times. We will use this approach, and, in every step, offer an 
arbitrary input pattern with activity $a_\text{I}$ to the net.

Since it may take a very long time before the distribution of all weights has 
reached its equilibrium by the effect of the Anti--Hebbian learning rule, we 
will not start with entirely arbitrary weights, but, instead, start from a rough 
approximation. We choose to initialize with weights that are distributed 
according to the Gaussian distribution (\ref{Fx}). For weights connecting 
neurons of the input layer to neurons of the hidden layer, the changes of the 
weights due to the Anti--Hebbian learning rule will be of the order of 
$\rho_\text{H}$. Therefore, we suppose that the width of the distribution of the 
weights will also be of the order of $\rho_\text{H}$, and we take
\begin{equation}\label{sigmah}
\textstyle \sigma = \frac{1}{2}\rho_\text{H}
\end{equation}
for the standard deviation of the initial Gaussian distribution. For the mean 
$\mu$ we will use a value $w_\text{H}$, which is a rough approximation of the 
average value of the strength of the weights connecting neurons of the input 
layer to neurons of the hidden layer. Since the Anti--Hebbian learning rule 
causes each neuron to change its state $x_i$ over and over again, the membrane 
potentials $h_i$ will fluctuate around the threshold potentials $\theta_i$. As 
an approximation we suppose that each membrane potential $h_i$ of a neuron of the 
hidden layer will be, on average, equal to its threshold potential:
\begin{equation}\label{hi0}
h_i \approx \theta_\text{H}, \quad i \in \text{H}.
\end{equation}
We now approximate $h_i$, given by (\ref{potential}), according to
\begin{equation}\label{hi1}
h_i \approx w_\text{H} \sum_{j \in V_i} x_j.
\end{equation}
The sum in this formula can be approximated by
\begin{equation}\label{hi2}
\sum_{j \in V_i} x_j \approx \overline{a_\text{I}}N_\text{I}(1 - d_\text{H}),
\end{equation}
with $\overline{a_\text{I}}$ the average activity of the input layer, which, in 
our model, will be equal to the average activity of the input patterns. 
Furthermore, $N_\text{I}(1 - d_\text{H})$ is an approximation for the number of 
presynaptic neurons. Substituting (\ref{hi2}) and (\ref{hi0}) into (\ref{hi1}), 
we obtain a rough approximation of the average value of the strengths of the 
weights $w_\text{H}$:
\begin{equation}\label{wijH}
w_\text{H} \approx \frac{\theta_\text{H}}{\overline{a_\text{I}}N_\text{I}(1 -
d_\text{H})}
\end{equation}

In the same way, we find for the weights connecting the neurons of the hidden 
layer to the neurons of the output layer
\begin{equation}\label{wij0}
w_\text{O} \approx \frac{\theta_\text{O}}{\overline{a_\text{H}}N_\text{H}(1 -
d_\text{O})}
\end{equation}
where $\overline{a_\text{H}}$ is the average activity of the hidden layer. As 
will turn out later, this average activity will be around the parameter 
$\alpha_\text{H}$ occurring in the Anti--Hebbian learning rule. We will use 
$\alpha_\text{H}$ as an estimation for $a_\text{H}$ in the above formula. 
Analogous with (\ref{sigmah}), $\frac{1}{2}\rho_\text{O}$ will be used for the 
standard deviation $\sigma$ for the weights connecting neurons of the hidden 
layer to neurons of the output layer.

\section{\label{num1}Effect of the Anti--Hebbian component}
It proves useful to study the Hebbian and Anti--Hebbian learning rules 
separately. Since the Hebbian rule has been studied earlier, we here consider 
Anti--Hebbian learning; in the next section the complete rule will be
discussed.

We will show that the activities, averaged in time, of the hidden and output 
layers are given by $\alpha_\text{H}$ and $\alpha_\text{O}$ respectively. 
Moreover, we will show that the distribution of the activities around the value 
$\alpha_\text{H}$ (or $\alpha_\text{O}$) is such that, effectively, each neuron 
$i$ in layer H (or O) has a probability $\alpha_\text{H}$ (or $\alpha_\text{O}$) 
to fire, independent of the activities of the other neurons. In other words, the 
proposed learning rule (\ref{deltawijA2}) focuses the average activity in a 
natural way around the values $\alpha_\text{H}$ and $\alpha_\text{O}$.

In case the threshold potentials $\theta_\text{H}$ and $\theta_\text{O}$ vanish, 
we will find that the state of a neuron in the output layer behaves independently of 
the states of the other neurons of the output layer. In case the threshold 
potentials do not vanish, we will observe strong correlations between the states of 
the output neurons, which vanish, however, when the dilution is taken to be 
sufficiently high.

\subsection{\label{secvanish}Vanishing threshold potentials}
We start by studying the case that $\theta_\text{H} = \theta_\text{O} = 0$ and 
$d_\text{H} = d_\text{O} = 0$. In order to get a first impression of the network 
behavior, we will plot the activity as a function of time, as well as, in a 
histogram, the distribution of its values ($0 \leq a \leq 1$). We will offer the 
network $p = 1000$ input patterns. Each input pattern will be repeatedly offered 
to the net until the output pattern to be associated with the input pattern is 
found. The input and associated output patterns are chosen at 
random but with a certain specified activity. As soon as the correct output is 
produced, the next pattern is presented until all output patterns associated with 
the input patterns have been found.

\begin{figure*}
\includegraphics*[width=\columnwidth]{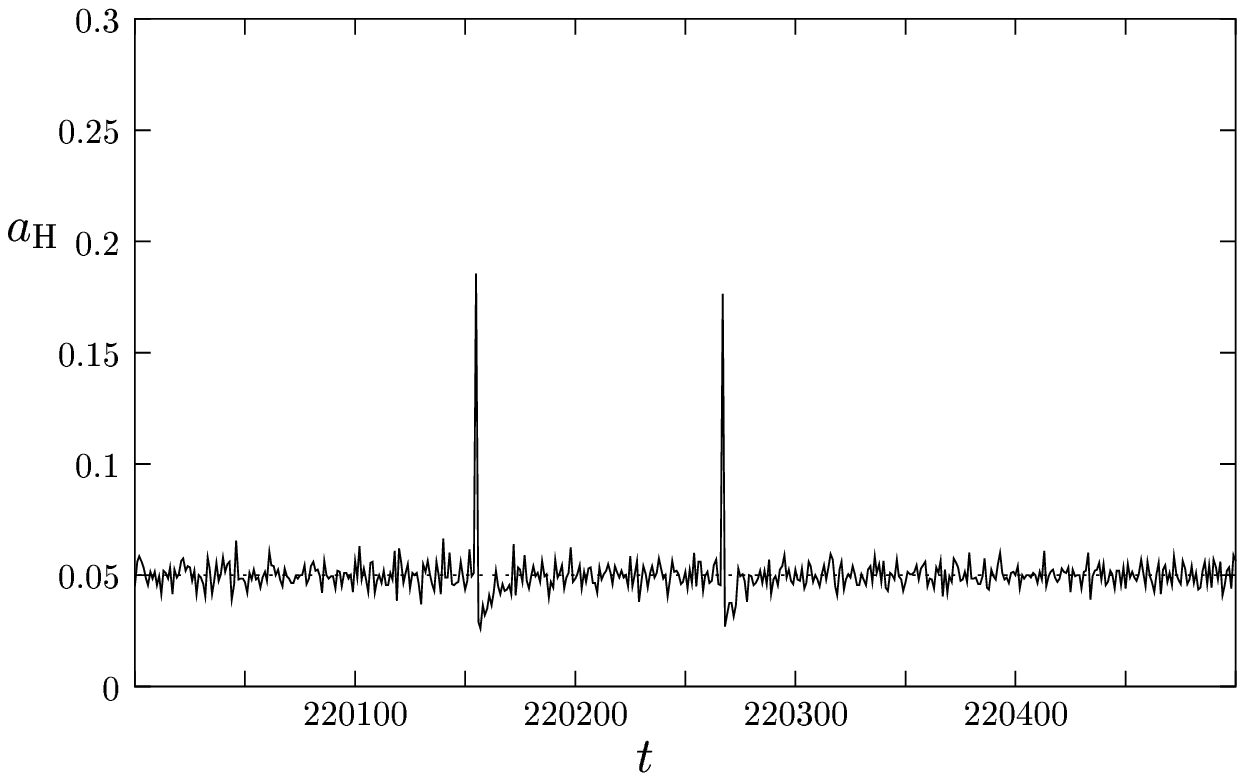}
\includegraphics*[width=\columnwidth]{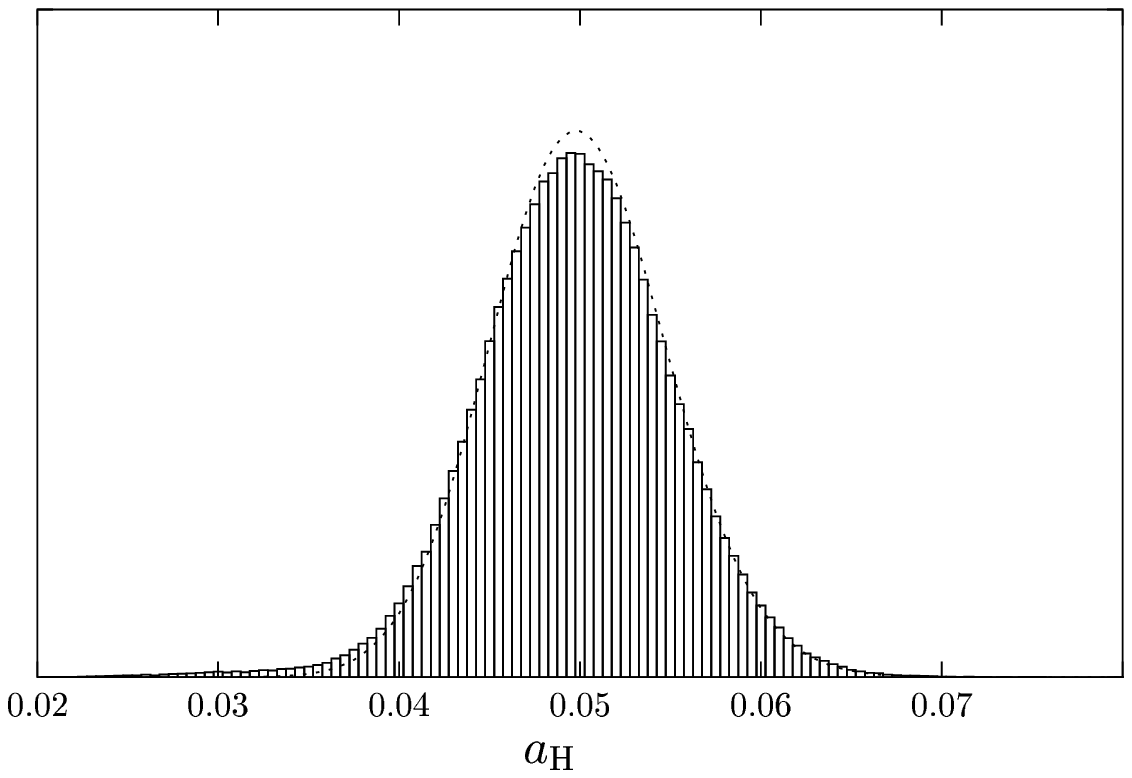}\\
\includegraphics*[width=\columnwidth]{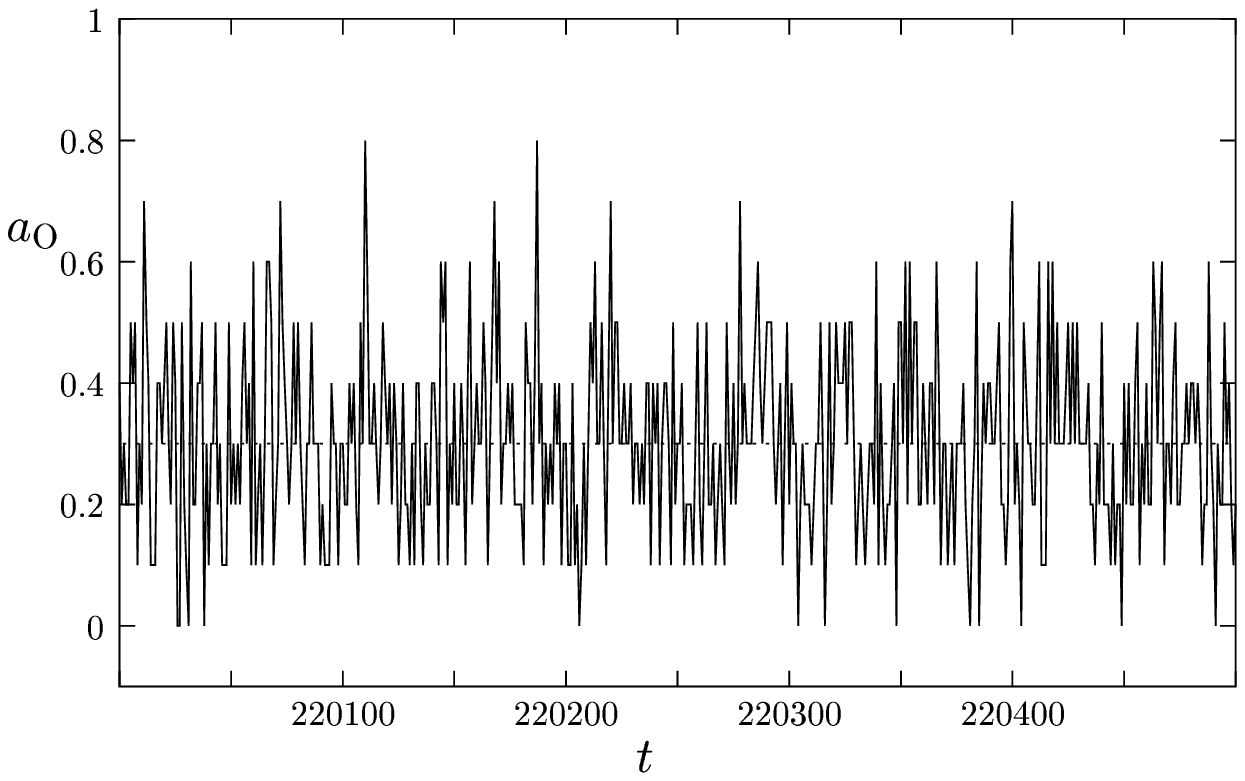}
\includegraphics*[width=\columnwidth]{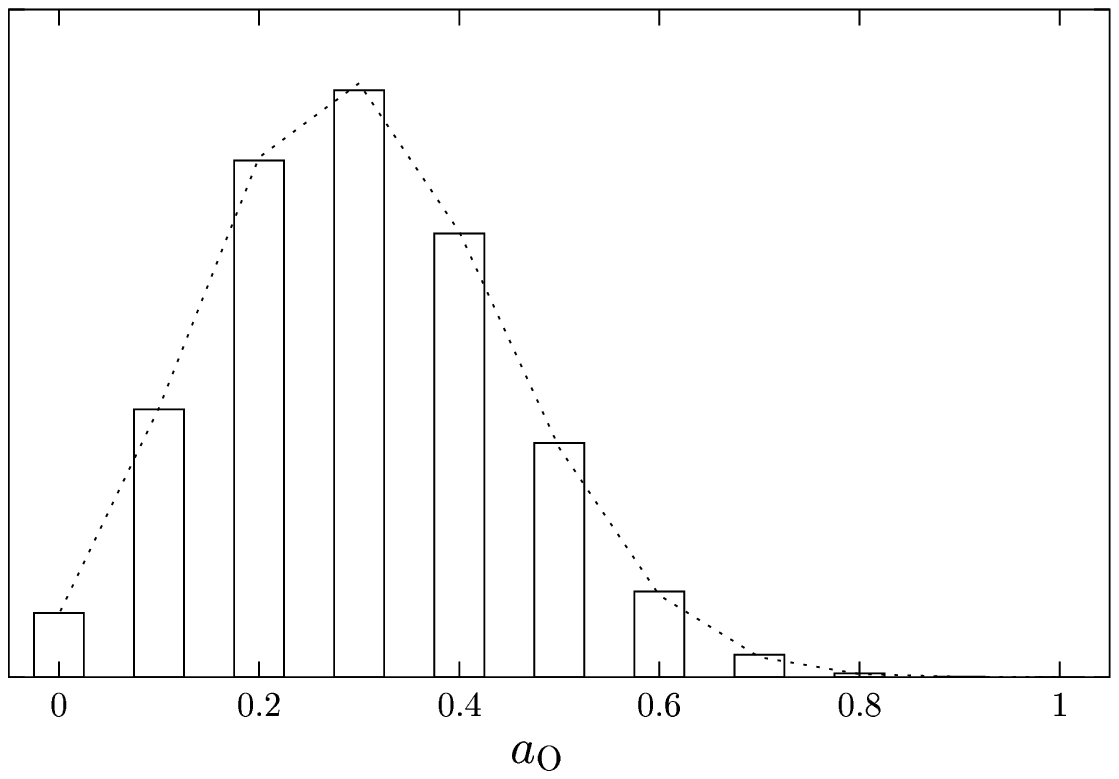}
\caption{The activities of the hidden layer (top) and the output layer (bottom) 
as a function of time (left), and their distributions (right). The network has 
$N_\text{I} = 20$ neurons in the input layer, $N_\text{H} = 2000$ neurons in the 
hidden layer and $N_\text{O} = 10$ neurons in the output layer. All neuron 
thresholds vanish: $\theta_\text{H} = \theta_\text{O} = 0$. The dilutions are 
zero: $d_\text{H} = d_\text{O} = 0$. The parameters in the learning rule are: 
learning rate $\eta = 0$, punishing rate $\rho = 0.01$, $\alpha_\text{H} = 0.05$ 
and $\alpha_\text{O} = 0.3$. The noise parameter is $\delta = 0.1$. The number 
of patterns is $p = 1000$. The number of active neurons in the input patterns was
$N_\text{I}^{(a)} = 3$.
The left pictures show only a small interval (500 
time steps) of the total number of 429,919 time steps needed to find all desired 
output patterns. The activities $a_\text{H}$ and $a_\text{O}$ are seen to wiggle 
around the values $\alpha_\text{H}$ and $\alpha_\text{O}$ (left pictures, dashed 
lines). The distribution of the activities (right pictures, bars) have a 
distribution resembling the distribution $Q(a)$ [eq.\ (\ref{Qa})] (right figures, 
dashed lines).}
\label{figa}
\end{figure*}
In figure \ref{figa} (top) we can observe that the activity $a_\text{H}$ of the 
hidden layer fluctuates around the value $\alpha_\text{H} = 0.05$, the value 
which we have chosen in the learning rule for changes of weights of connections 
between the input and hidden layers. This is what we had hoped to achieve when 
we postulated (\ref{deltawijA2}). The larger fluctuations occur when the net is 
confronted with a new input pattern, to which it must learn to react by a new, 
prescribed output pattern. We see that it takes only a short period of time 
before the net has found back its balance. Similar observations can be made with 
respect to the output layer: in figure \ref{figa} (bottom) the activity is seen 
to fluctuate around the value $\alpha_\text{O} = 0.3$, chosen in the learning 
rule (\ref{deltawijA2}). It can also be seen that the distributions closely 
resemble the distribution $Q(a)$ [eq.\ (\ref{Qa})]. This means that the 
Anti--Hebbian learning rule effectively causes each neuron to behave and fire, 
with a probability $a_\text{X}$, as if it is independent of the states of the 
other neurons.

\subsection{\label{nonvan}Non--vanishing threshold potentials}
\begin{figure*}
\includegraphics*[width=\columnwidth]{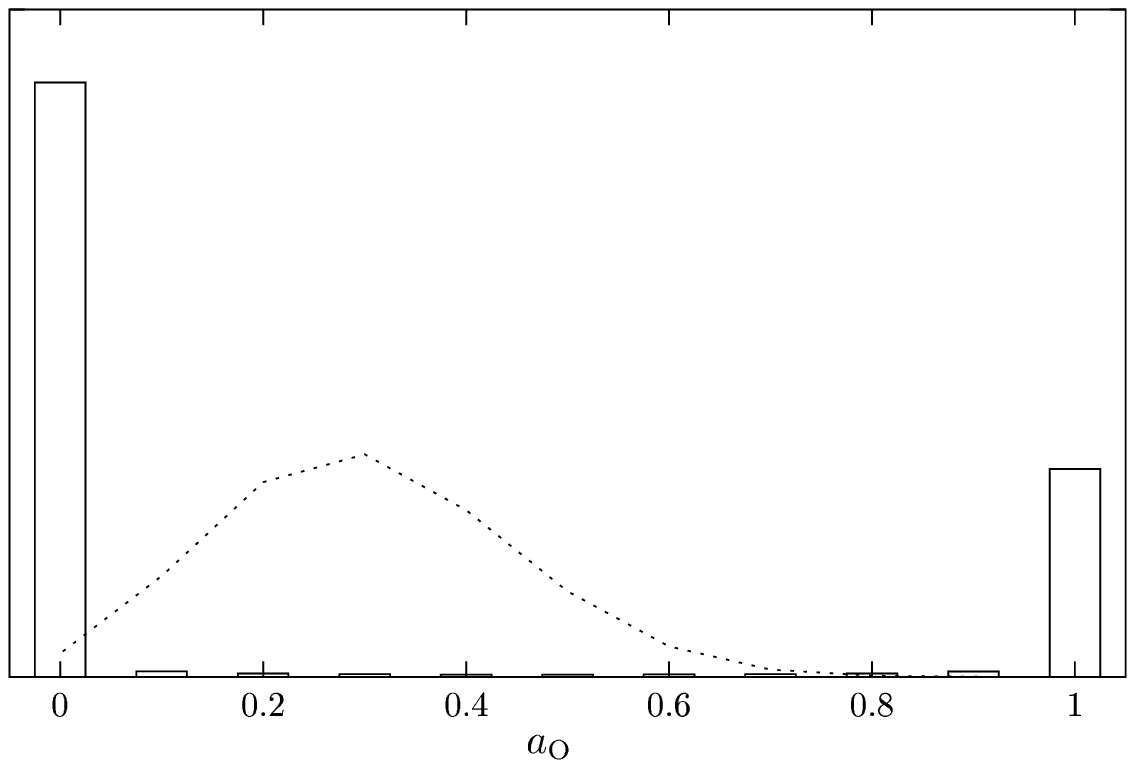}
\includegraphics*[width=\columnwidth]{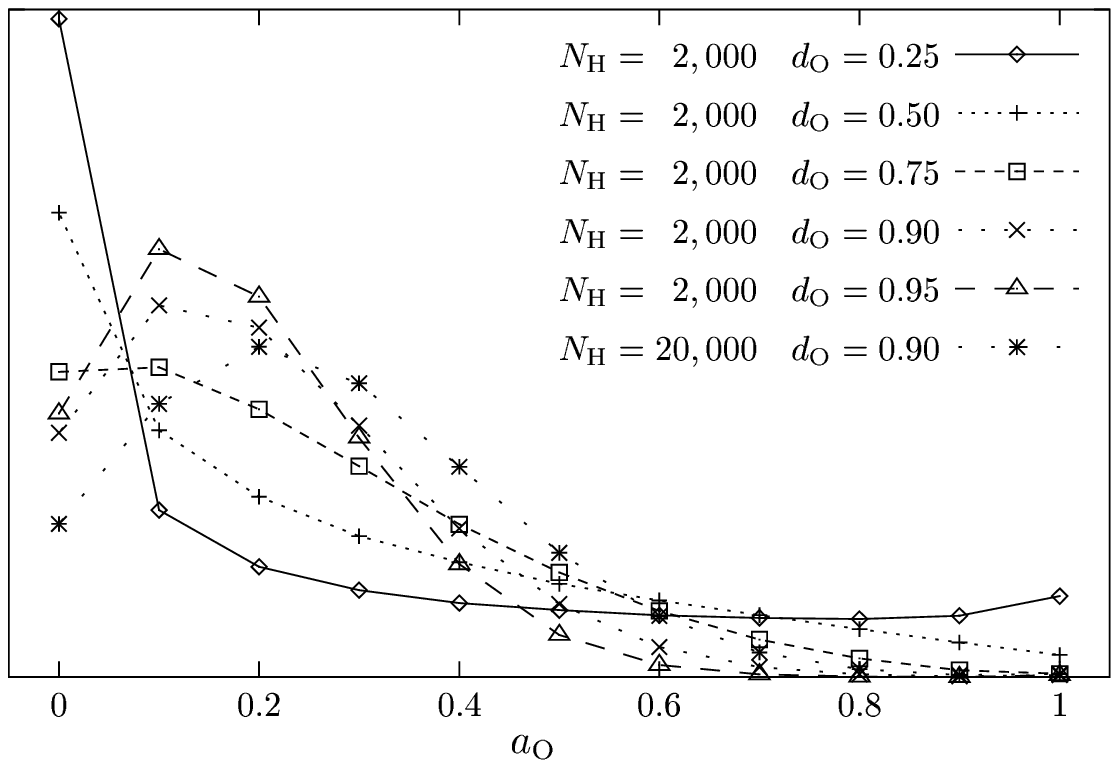}
\caption{The distribution of the activity of the output layer for 
$\theta_\text{H} = \theta_\text{O} = 1$ and $d_\text{H} = 0$; other 
parameters are taken the same as in figure \ref{figa}. The neurons of the output 
layer are seen to fire strongly correlated in case $d_\text{O} = 0$ (left, bars): they 
almost always either all are active ($a_\text{O} = 1$) or 
quiescent ($a_\text{O} = 0$). If no 
correlations at all would be present, the dashed line of the left figure would 
be found for the distribution of the neuron activity. This dashed line is given 
analytically by $Q(a)$ [eq.\ (\ref{Qa})]. 
When $d_\text{O}$ increases (right), the correlations of the activities of
the output neurons are seen to decrease.
Around the value $d_\text{O} = 0.9$ the activity
distribution resembles the distribution $Q(a)$
the most. The resemblance becomes better if the number of hidden 
neurons, $N_\text{H}$, increases.}
\label{d}
\end{figure*}
In this section we study a network that is the same as in the preceding section, 
but with non--vanishing threshold potentials: we will take $\theta_\text{H} = 
\theta_\text{O} = 1$. When we now repeat the simulations, we find one important 
difference: the neurons of the output layer turn out to exhibit strong 
correlations: almost always, either all of them are active or all of them are 
non--active. This can be seen in figure \ref{d} (left). This is alarming and 
most undesirable, since it would imply that the network has, effectively, only 
two different output states. Note, that although the probability that 
$a_\text{O}$ equals $\alpha_\text{O}$ now is very small, 
its average $\overline{a_\text{O}}$ is still close to $\alpha_\text{O}$.

It can be expected that if the average number of neurons of some layer
$\text{X}_B$ that are postsynaptic to a neuron of a preceding layer
$\text{X}_A$ becomes lower, the correlated behavior of the neurons of
$\text{X}_B$, which, apparently, was induced by the neurons of $\text{X}_A$,
will decrease.
In order to verify this, we now study a network with 
non--vanishing dilution. In figure \ref{d} (right), we plotted the distribution 
of the activity $a_\text{O}$ of the output layer for five different values of the 
dilution $d_\text{O}$. Comparing the left and right pictures of figure \ref{d} 
we observe a dramatic change in the behavior indeed: the undesired effect 
quickly diminishes with increasing dilution.

Note that the distribution is shifted somewhat towards zero in comparison with 
the distribution of uncorrelated output neurons. The shift decreases when the
number of neurons of the hidden layer increases: for $N_\text{H} = 20,000$ there
is a better resemblance to the $Q(a)$--curve than for $N_\text{H} = 2,000$.
For non--zero dilution the probability that none of the neurons that
are presynaptic to an output neuron fire increases, resulting in a lower 
average activity. This causes the shift of the activity distribution to lower
values.

\section{\label{num2}Testing the full learning rule}
In the preceding section we saw that the Anti--Hebbian part of the learning rule 
adapts the weights in such a way that a desired input--output relation is found 
after a number of time steps, while, at the same time the network activities 
$a_\text{H}$ and $a_\text{O}$ stay within acceptable bounds.
In this section we study the complete learning rule (\ref{j1wij}) with
(\ref{deltawijH2}) and (\ref{deltawijA2}).

\subsection{\label{seccomp}Comparing to other models}
\begin{figure*}
\includegraphics*[width=\columnwidth]{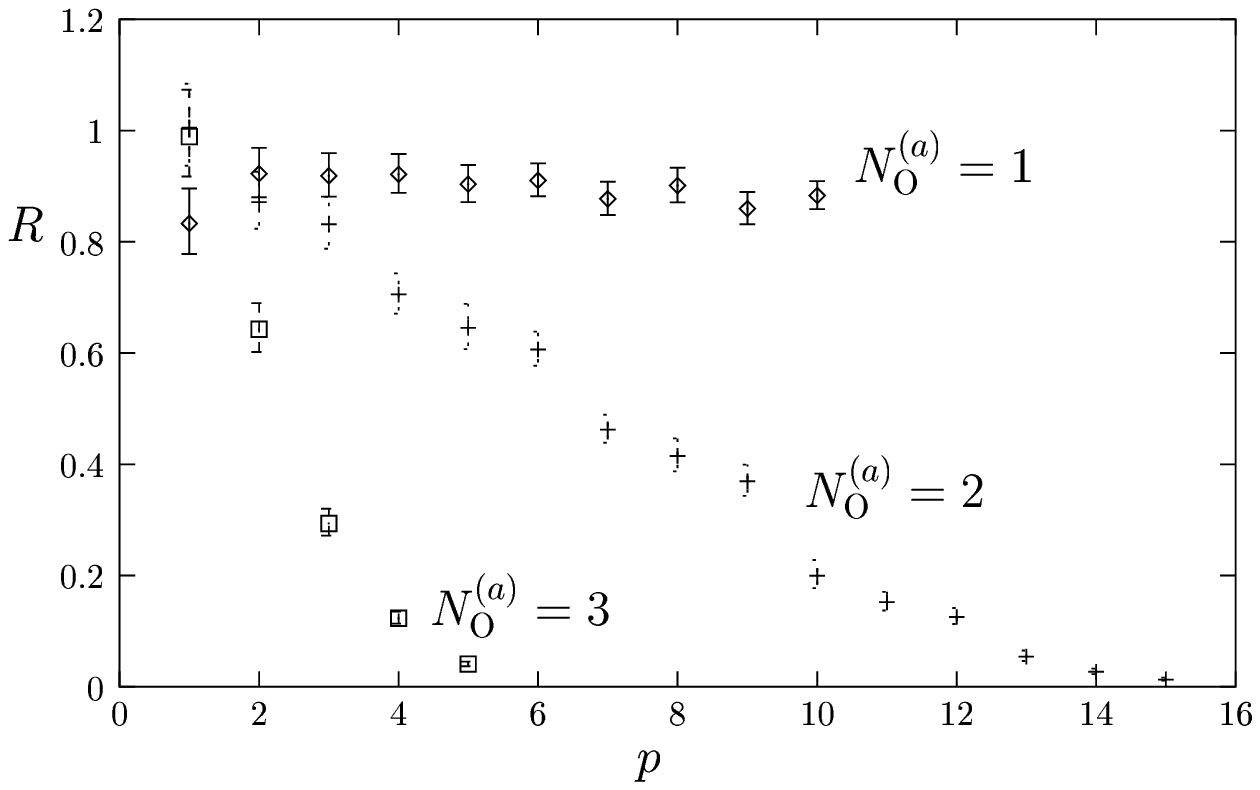}
\includegraphics*[width=\columnwidth]{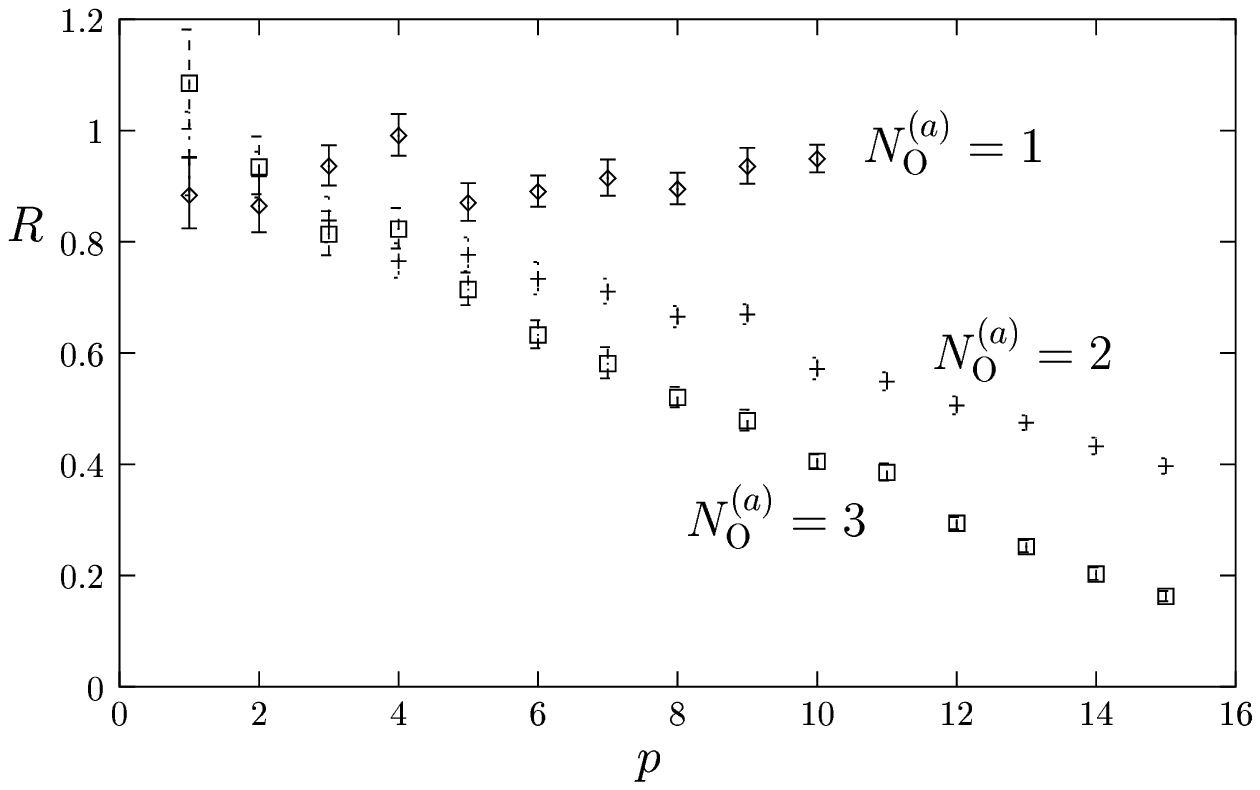}\\
\includegraphics*[width=\columnwidth]{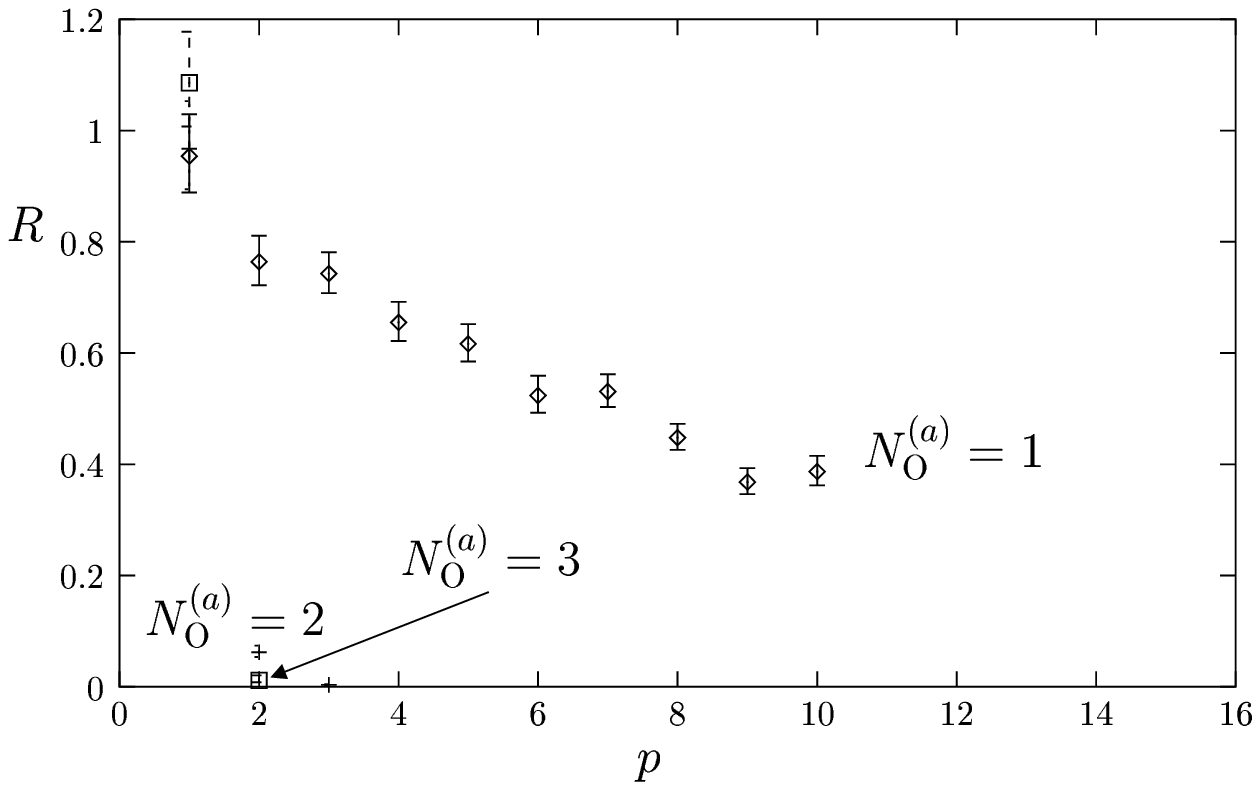}
\includegraphics*[width=\columnwidth]{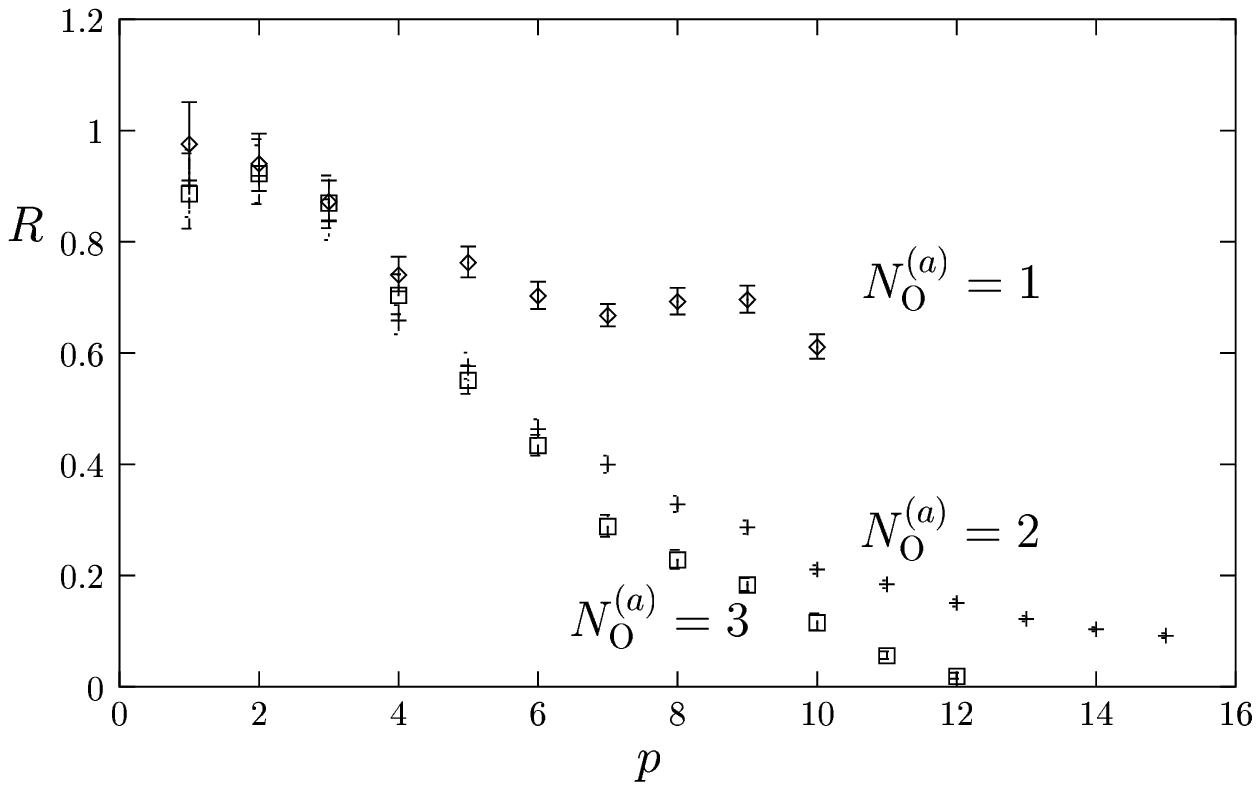}
\caption{Comparing realistic (bottom) and extremal dynamics (top) with (right) 
and without (left) a rewarding component in the learning rule. The performance 
is measured as a function of the number of input--output relations $p$. We have 
chosen $N_\text{I} = N_\text{O} = 10$, $N_\text{H} = 2000$, $N_\text{I}^{(a)} = 
N_\text{O}^{(a)} = 1$, 2, or 3, $\theta_\text{H} = \theta_\text{O} = 0$, 
$d_\text{H} = d_\text{O} = 0$, $\rho = 0.01$, $\eta = 0$ (left pictures), $\eta 
= 0.02$ (right pictures), $\alpha_\text{H} = 0.025$, 
$\alpha_\text{O} = N_\text{O}^{(a)} / N_\text{O}$ and $\delta = 0.1$.}
\label{explore}
\end{figure*}

In order to make contact with existing literature, we also simulate systems with 
extremal dynamics. We consider input--output relations with the same number 
of active neurons $N_\text{I}^{(a)} = N_\text{O}^{(a)} = N^{(a)}$ 
in the input and output patterns. In figure 
\ref{explore} we plotted the performance $R$ of a network as a function of the 
number of input--output relations $p$ to be learned, for $N^{(a)} = 1$, $N^{(a)} 
= 2$ and $N^{(a)} = 3$. The two pictures at the top are for the case of extremal 
dynamics, the pictures at the bottom correspond to simulations for the more 
realistic case that the neurons fire if their membrane potentials exceed their 
thresholds. The two pictures at the left are without reward ($\eta = 0$), those 
at the right are with reward ($\eta \neq 0$).

We conclude that the net is able to learn a number of input--output relations
indeed.
The situation $N^{(a)} = 1$ (top, left) is analogous to the case considered by 
Chialvo and Bak. 
In the simple case that $N^{(a)} = 1$, the performance is almost perfect, i.e., 
$R$ is close to 1. However, the performance decreases quickly for the case that 
more than two neurons are active in the input and output patterns. In figure 
\ref{explore} (top, right) we observe that rewarding significantly increases the 
performance $R$ if $N^{(a)}$ is larger than 1. A similar observation has been 
made earlier by Bosman et al.\ \cite{bosman}.

Figure \ref{explore} (left, bottom) is to be compared with figure \ref{explore} 
(left, top). For $N^{(a)} = 1$, the performance is not as good as in case of 
extremal dynamics, but it still works satisfactory. However, for values of 
$N^{(a)}$ larger than one, the performance is very bad. Adding reward to the 
learning rule, we find the results of figure \ref{explore} (right, bottom). The 
improvement of the performance is impressive. Hence, our model performs the task 
of realizing prescribed input--output relations reasonably well, although its 
performance is not as good as in the less realistic case that extremal dynamics 
is used.

We close this section with the following observation: In case only one 
input--output relation is to be learned by the net, i.e., $p = 1$, the performance 
$R$ is close to 1, as can be read off from figure \ref{explore}. This implies 
that the period of search for the correct output by means of the Anti--Hebbian 
term is close to the random search time. In other words, in case the feed--back
is binary only ($r = 0$ or $r = 1$), Anti--Hebbian learning enables a way of
search which is close to optimal.

\subsection{\label{varying}Numerical experiments on the influence of parameters}
The performance of a neural net depends on many parameters, e.g., the 
coefficients $\eta$, $\rho$ and $\alpha$ occurring in the learning rule, 
the dilution $d$ and the threshold potential $\theta$. In 
this section we take the thresholds $\theta_\text{X} = 0$ 
($\text{X} = \text{I}, \text{H}, \text{O}$). In order to get more 
insight regarding the behavior of the net in dependence of variations of all 
these parameters, we will study two particular cases.

\begin{figure}
\includegraphics*[width=\columnwidth]{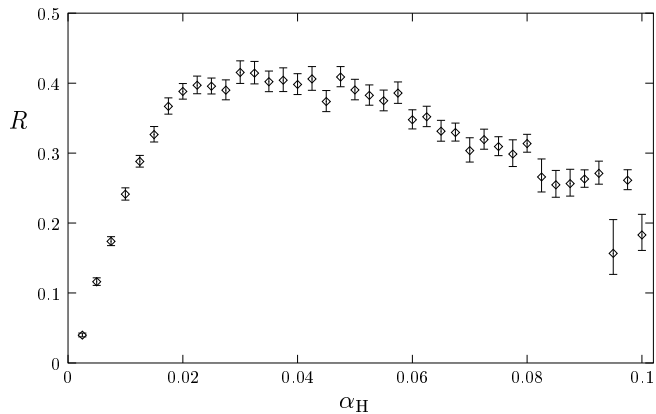}
\caption{Performance $R$ as a function of the parameter $\alpha_\text{H}$ 
determining the activity of the hidden layer for $p = 10$ input--output 
relations. We set $N_\text{I} = N_\text{O} = 10$, $N_\text{H} = 2000$, 
$N_\text{I}^{(a)} = N_\text{O}^{(a)} = 2$, $\theta_\text{H} = \theta_\text{O} = 
0$, $d_\text{H} = d_\text{O} = 0$, $\rho = 0.1$, $\eta = 0.2$, $\delta = 0.1$. 
The performance is best for values of $\alpha_\text{H}$ between $0.02$
and $0.06$.}
\label{alphaH}
\end{figure}
In figure \ref{alphaH} we plotted the performance $R$ of the net as a function 
of the parameter $\alpha_\text{H}$, the coefficient occurring in the 
Anti--Hebbian part of the learning rule, which determines the activity of the 
hidden layer. There is seen to be an optimum in the performance. For values of 
$\alpha_\text{H}$ below $0.02$, the performance decreases rapidly. It follows 
that the parameter $\alpha_\text{H}$ occurring in the Anti--Hebbian part of the 
learning rule, a neuron property, has an important effect on the performance of 
the net. It is seen that the performance diminishes when the activity becomes 
too large, in agreement with earlier results \cite{bak}.

\begin{figure}
\includegraphics*[width=\columnwidth]{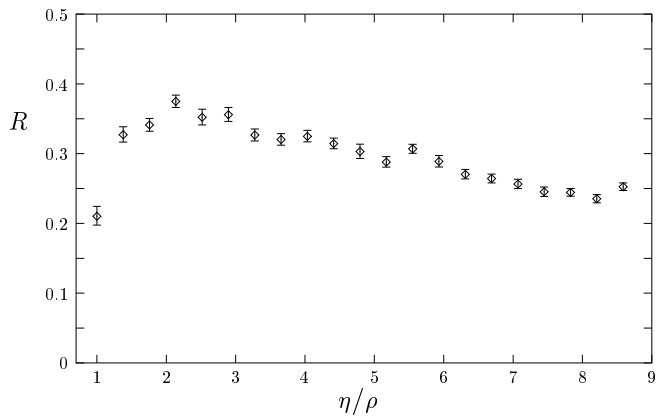}
\caption{Performance as a function of the ratio of reward and punishment for $p 
= 10$ input--output relations. We chose $\alpha_\text{H} = 0.025$, $\rho = 0.05$
and $\eta$ was varied between 
$0.05$ and $0.5$. Other parameters where chosen the same as in the 
simulations of figure \ref{alphaH}. 
For values of $\eta/\rho$ around 2, the performance seems 
optimal.}
\label{etarho}
\end{figure}
In figure \ref{etarho} we plotted the performance $R$ as a function of 
$\eta/\rho$ for $\rho = 0.05$. For values of $\eta$ around two times the value 
of $\rho$, the performance is optimal. We conclude that when the rewarding part 
of the learning rule (proportional to $\eta$) becomes smaller and smaller, the 
performance of the net strongly decreases. On the other hand, if the rewarding 
part becomes larger and larger, the performance also diminishes, albeit more 
slowly. These observations do not come as a surprise: evidently, if the 
memorizing Hebbian term is too weak -- relative to the Anti--Hebbian term -- 
learning will be slow, whereas it will also be slow if the Hebbian term
becomes too strong, since then the learning of any pattern will change the 
existing connections too wildly.

\subsection{\label{nonvan2}Non--vanishing threshold potentials}
\begin{figure}
\includegraphics*[width=\columnwidth]{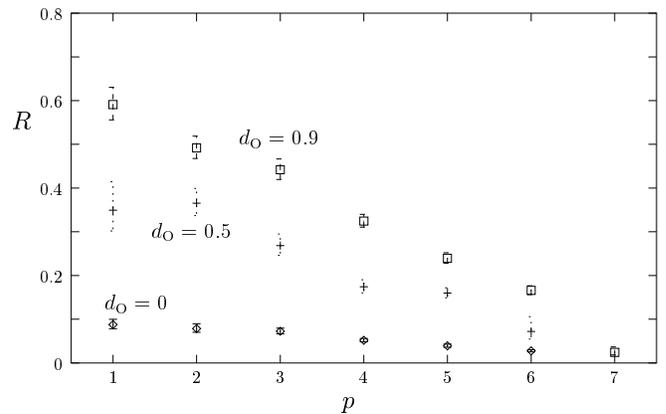}
\caption{Performance of a network with $\theta_\text{H} = \theta_\text{O} = 1$ 
as a function of the number of input--output relations $p$ for different values 
of the dilution $d_\text{O}$: $d_\text{O} = 0$, $d_\text{O} = 0.5$, $d_\text{O} 
= 0.9$. We chose 
$\alpha_\text{H} = 0.025$. Other parameters 
are the same as those in figure \ref{alphaH}. The network performance is 
best in this picture for $d_\text{O} = 0.9$.}
\label{thetad}
\end{figure}
In view of section \ref{nonvan} we expect that also in case of use of
the full learning rule (\ref{j1wij}) with
(\ref{deltawijH2}) and (\ref{deltawijA2})
we will only get satisfactory results if we take the dilution $d$ unequal 
to zero. In figure \ref{thetad} the performance is plotted as a function of the 
number of input--output relations $p$ for different values of the dilution 
$d_\text{O}$. The network performance is optimal around $d_\text{O} = 0.9$. The 
fact that a dilution here actually enhances the performance is a consequence of 
the fact that undesired correlations (see section \ref{nonvan}) diminish with 
increasing dilution.

\section{\label{conclusions}Conclusions}
Our goal to model, in a biologically plausible way, neural networks capable of 
learning without the use of extremal dynamics 
\cite{bak} or some other mechanism to control the neuronal activity was reached 
successfully.

In section \ref{hebbvsantihebb} we found, on the basis of four biological 
assumptions, two possible learning rules,
given by the equations (\ref{H}) and (\ref{A}). By analyzing 
their effects, we were able to associate them in a unique way with reward and 
punishment, and to formulate a plausible learning rule (\ref{j1wij}). For the 
rewarding part, we used a form derived in \cite{leeuwen}. for the punishing part 
we chose two constants [see eq.\ (\ref{epsilonijAprime})]. By studying the 
effects of this specific form with two constants, 
we found that this punishing part of our learning 
rule is able to control the average activity (average fraction of firing 
neurons) in a neuronal net. We showed that the activity remained around a desired 
level while, at the same time, the network is searching for the correct output by 
generating random patterns with a desired activity with great efficiency.

Finally, we showed that for neurons with non--zero threshold potentials, the 
neural net must be diluted in order to achieve a reasonable performance: 
dilution is found to enhance the functionality of a biological neural net. This 
is because the neurons start to behave correlated when the threshold potentials 
are non--zero. When the dilution increases, these undesired correlations 
decrease. In nature neural nets normally are strongly diluted: the human brain, 
for example, has of the order of $10^{11}$ neurons but there are only 
$10^4$--$10^5$ synapses per neuron.

\section{\label{outlook}Outlook}
Evidently, the study on the behavior of biologically inspired neural networks is 
far from complete.

One may, for instance, include more biologically known features into the model 
like time delay of the axonal signal or the refractory period of a neuron.
Also, the architecture of the net could 
be made more realistic by including more layers and adding feed--back and 
lateral connections. Furthermore, the model of the surrounding world, i.e., the 
input--output relations to be learned could be made more realistic, for example 
by using a dynamic model of a realistically changing world. Also, the neuron 
model could be made to resemble closer real neurons by including different types 
of neurons and synapses, or using the fact that excitatory synapses are probably 
more plastic than the inhibitory ones. Another extension would be to refine the 
measure of success. Instead of just a binary feed--back signal indicating 
whether the output is \emph{right} or \emph{wrong} in reaction to some input, a 
feed--back signal that can take on a range of values is possible. Also, a 
relative measure could be used, telling the network if it has performed better 
or worse than during a previous attempt.

A most obvious first extension of our particular model would be to include 
(inhibitory) lateral connections inside layers and possibly also feed--back 
connections. It would be interesting to study the influence of these 
non--feed--forward connections on the behavior of the network, and, especially, 
the effect they will have on the (average) activity of the network. Another 
extension of our model could be to associate Hebbian and Anti--Hebbian learning 
with different types of neurons (or synapses) instead of letting the same neurons 
behave differently under different conditions of success or failure. Such a 
model could have the advantage that the Hebbian connections, in which the 
input--output relations are memorized on success, would not have to be changed on 
failure. Thus, instead of changing the same connections, different, 
Anti--Hebbian connections could possibly do the job of searching for successful 
output patterns, without changing the already learned input--output relations.

As a final remark, we mention that different extensions or alterations are 
possible to enhance the performance of the network, which may, however, be 
implausible biologically. For example, if an input--output relation would not be 
strengthened over and over again once it has been learned, the network generally 
would perform better. 

\begin{acknowledgments}
The authors are indebted to Bastian Wemmenhove for a number of very useful
suggestions.
\end{acknowledgments}

\end{document}